\documentstyle[aps,preprint,pra]{revtex}
\begin{document}
\newcommand{\etal}{{\em et al.}\/}
\newcommand{\IP}{inner polarization}
\newcommand{\IPF}{\IP\ function}
\newcommand{\IPFs}{\IP\ functions}
\newcommand{\auth}[2]{#1 #2, }
\newcommand{\jcite}[4]{#1 #2 (#4) #3}
\newcommand{\et}{ and }
\newcommand{\erratum}[3]{\jcite{erratum}{#1}{#2}{#3}}
\newcommand{\twoauth}[4]{#1 #2 and #3 #4,}
\newcommand{\oneauth}[2]{#1 #2,}
\newcommand{\andauth}[2]{and #1 #2, }
\newcommand{\book}[4]{{\it #1} (#2, #3, #4)}
\newcommand{\JCP}[3]{\jcite{J. Chem. Phys.}{#1}{#2}{#3}}
\newcommand{\jms}[3]{\jcite{J. Mol. Spectrosc.}{#1}{#2}{#3}}
\newcommand{\jmstr}[3]{\jcite{J. Mol. Struct.}{#1}{#2}{#3}}
\newcommand{\cpl}[3]{\jcite{Chem. Phys. Lett.}{#1}{#2}{#3}}
\newcommand{\cp}[3]{\jcite{Chem. Phys.}{#1}{#2}{#3}}
\newcommand{\pr}[3]{\jcite{Phys. Rev.}{#1}{#2}{#3}}
\newcommand{\jpc}[3]{\jcite{J. Phys. Chem.}{#1}{#2}{#3}}
\newcommand{\jpca}[3]{\jcite{J. Phys. Chem. A}{#1}{#2}{#3}}
\newcommand{\jpcA}[3]{\jcite{J. Phys. Chem. A}{#1}{#2}{#3}}
\newcommand{\jpcB}[3]{\jcite{J. Phys. Chem. B}{#1}{#2}{#3}}
\newcommand{\jcc}[3]{\jcite{J. Comput. Chem.}{#1}{#2}{#3}}
\newcommand{\molphys}[3]{\jcite{Mol. Phys.}{#1}{#2}{#3}}
\newcommand{\cpc}[3]{\jcite{Comput. Phys. Commun.}{#1}{#2}{#3}}
\newcommand{\jcsfii}[3]{\jcite{J. Chem. Soc. Faraday Trans. II}{#1}{#2}{#3}}
\newcommand{\prsa}[3]{\jcite{Proc. Royal Soc. A}{#1}{#2}{#3}}
\newcommand{\jacs}[3]{\jcite{J. Am. Chem. Soc.}{#1}{#2}{#3}}
\newcommand{\ijqcs}[3]{\jcite{Int. J. Quantum Chem. Symp.}{#1}{#2}{#3}}
\newcommand{\ijqc}[3]{\jcite{Int. J. Quantum Chem.}{#1}{#2}{#3}}
\newcommand{\spa}[3]{\jcite{Spectrochim. Acta A}{#1}{#2}{#3}}
\newcommand{\tca}[3]{\jcite{Theor. Chem. Acc.}{#1}{#2}{#3}}
\newcommand{\tcaold}[3]{\jcite{Theor. Chim. Acta}{#1}{#2}{#3}}
\newcommand{\jpcrd}[3]{\jcite{J. Phys. Chem. Ref. Data}{#1}{#2}{#3}}

\draft
\title{The heat of atomization of sulfur trioxide, SO$_3$ --- a 
benchmark for computational thermochemistry}
\author{Jan M.L. Martin}
\address{Department of Organic Chemistry,
Kimmelman Building, Room 262,
Weizmann Institute of Science,
IL-76100 Re\d{h}ovot, Israel. {\rm E-mail:} {\tt comartin@wicc.weizmann.ac.il}
}

\date{{\em Chemical Physics Letters}, in press (received June 3, 1999)}
\maketitle
\begin{abstract}
Calibration ab initio (direct coupled cluster) calculations
including basis set extrapolation, relativistic effects, inner-shell
correlation, and an anharmonic zero-point energy,
predict the total atomization energy at 0 K of SO$_3$ to be 335.96
(observed 335.92$\pm$0.19) kcal/mol. 
Inner polarization functions make very large (40 kcal/mol with
$spd$, 10 kcal/mol with $spdfg$ basis sets) contributions to the SCF
part of the binding energy. The molecule presents an unusual hurdle
for less computationally intensive theoretical thermochemistry methods
and is proposed as a benchmark for them. A slight modification of
Weizmann-1 (W1) theory is proposed that appears to significantly
improve performance for second-row compounds.
\end{abstract}

\section{Introduction}

Neither the sulfuric anhydride (SO$_3$) molecule, nor its importance
in atmospheric and industrial chemistry, require any introduction
to the chemist.

SO$_3$ displays somewhat unusual bonding. While it is often cited
as a `hypervalent molecule' in undergraduate inorganic chemistry
textbooks, quantitative 
theories of chemical bonding such as atoms-in-molecules\cite{bader}
unequivocally
show (see Ref.\cite{Cio93} for a lucid review and discussion) that
there are no grounds for invoking violation of the octet rule in
SO$_3$ (or, for that matter, most second-row molecules), and that
bonding in SO$_3$ is best seen as a combination of moderately
polar $\sigma$ bonds with highly polar $p_{\pi,S},p_{\pi,O}$ bonds.

Previous experience on BF$_3$\cite{bf3} and SiF$_4$\cite{sif4} suggests
that in molecules with several strong and very polar bonds, basis set
convergence will be particularly slow. In addition, in a recent
calibration study on the anharmonic force field of SO$_3$ it was
found that the molecule represented a fairly extreme example
of a phenomenon noted previously for second-row molecules\cite{sio,so2,Bau95}
--- namely the great sensitivity of the SCF part of computed 
properties to the presence of so-called `inner polarization functions', 
i.e. high-exponent $d$ and $f$ functions. 

Very recently, Martin and de Oliveira\cite{w1} published a standard protocol
known as W2 (Weizmann-2) theory that was able to predict total atomization
energies of a fairly wide
variety of molecules (including SO$_2$, which is relevant
for this work) to better than 0.23 kcal/mol on average (0.18 kcal/mol for
molecules dominated by a single reference configuration). Application of
this method to SO$_3$ requires a CCSD (coupled cluster with all 
single and double excitations\cite{Pur82}) calculation with 529 basis
functions in the $C_{2v}$ nondegenerate subgroup, which was well beyond our 
available computational resources, particularly in terms of disk space.

Very recently, however, Sch\"utz et al.\cite{dirccsd} developed a
general implementation of integral-direct correlated methods
that made possible, inter alia, CCSD calculations on basis sets this
size on workstation computers. 
Consequently, we carried out a benchmark calculation on the
heat of atomization of SO$_3$, which is reported in the present work.

Having obtained the benchmark ab initio value, we will assess the 
performance of some less computationally demanding schemes. This 
includes W1 theory\cite{w1}, which is much more cost-effective than 
W2 theory but performs much less well for second-row than for first-row
compounds. From an analysis of the SO$_3$ results, we will derive
a minor modification (denoted W1$'$ theory) which in effect largely
removes this disadvantage.

\section{Methods}

Most electronic structure calculations were carried out using
MOLPRO98.1\cite{molpro}
(with integral-direct code\cite{dirccsd} installed)
running on a DEC Alpha 500/500 workstation
at the Weizmann Institute of Science. Some additional
calculations were carried out using GAUSSIAN 98\cite{g98} running on
the same platform.

As in our previous work on SO$_2$\cite{so2}, the CCSD(T) 
electron correlation method\cite{Rag89,Wat93}, as implemented 
by Hampel \etal\cite{Ham92}, has been used
throughout. The acronym stands for coupled cluster with all single
and double substitutions\cite{Pur82} augmented by a quasiperturbative 
account for triple excitations\cite{Rag89}. From extensive 
studies (see \cite{Lee95} for a review) this method is known to yield
correlation energies very close to the exact $n$-particle solution 
within the given basis set as long as the Hartree-Fock determinant is
a reasonably good zero-order reference wave function.
None of the usual indicators (${\cal T}_1$ diagnostic\cite{Lee89},
largest excitation amplitudes, or natural orbital occupancies of
first few HOMOs and LUMOs) suggest a significant departure from the
single-reference regime. (For the record, ${\cal T}_1$=0.018 for SO$_3$.)

Valence correlation basis sets are built upon the augmented 
correlation-consistent polarized $n$-tuple zeta (aug-cc-pV$n$Z,
or AV$n$Z for short)
basis sets of Dunning and coworkers\cite{Dun89,ccecc}.
In this work, we have considered AVDZ, AVTZ, AVQZ, and AV5Z
basis sets, with maximum angular momenta $l$=2 ($d$), 3 ($f$), 4 ($g$), 
and 5 ($h$), respectively. The effect of inner polarization was
accounted for by adding `tight' (high-exponent) $d$ and $f$ 
functions with exponents that follow even-tempered series $\alpha\beta^n$,
with $\alpha$ the tightest exponent of that angular momentum in the
underlying basis set and $\beta$=2.5. Such basis sets are denoted
AV$n$Z+d, AV$n$Z+2d, and AV$n$Z+2d1f. The largest basis set considered
in the present work, AV5Z+2d1f, corresponds to 
$[8s7p7d5f3g2h]$ on sulfur and $[7s6p5d4f3g2h]$ on oxygen
(148 and 127 contracted basis functions, respectively), adding up to
529 basis functions for the entire molecule. The CCSD calculation in
this basis set was carried out using the newly implemented\cite{dirccsd}
direct algorithm; all other CCSD and CCSD(T) calculations were done
conventionally.

The effect of inner-shell correlation was considered at the 
CCSD(T) level using two specialized core correlation basis sets,
namely the Martin-Taylor (MT) basis set\cite{hf} used in previous
work on SO$_2$\cite{so2}, and the
somewhat more compact MTsmall basis set that is used in the W2 protocol\cite{w1}
for this purpose. Correlation from the sulfur ($1s$) orbital was not
considered, since this lies too deep to meaningfully interact with the
valence orbitals. Scalar relativistic effects were computed as expectation
values of the first-order Darwin and mass-velocity corrections\cite{Cow76,Mar83}
for the ACPF (averaged coupled pair functional\cite{Gda88}) wave function
with the abovementioned core correlation basis sets. (All electrons
were correlated in these calculations since relativistic effects are
most important for the electrons closest to the nucleus.)

The CCSD(T)/VQZ+1 
reference geometry used throughout this work, 
$r_{SO}$=1.42279 \AA, was taken from the earlier
spectroscopic work on SO$_3$\cite{so3}, as was the anharmonic zero-point 
energy of 7.794 kcal/mol.

\section{Results and discussion}

The most striking feature of the basis set convergence at the SCF level
(Table 1) is certainly the great importance
of inner polarization functions: augmenting the AVDZ basis set with two
tight functions on S has an effect of no less than 40.5 kcal/mol!
The same operation affects the AVTZ SCF binding energy by 15.7
kcal/mol, and even from AVQZ to AVQZ+2d the effect is still 8.6 kcal/mol,
probably the largest such effect hitherto observed. In addition 
augmenting the basis set by a tight $f$ function has an effect of 1.1
kcal/mol from AVTZ+2d to AVTZ+2d1f, but only 0.16 kcal/mol from
AVQZ+2d to AVQZ+2d1f. Presumably the effect from AV5Z+2d to AV5Z+2d1f
will be next to negligible. 

Not surprisingly, this translates into a substantial effect on the
extrapolated SCF limit. A geometric extrapolation\cite{Fel92} 
from the AV\{D,T,Q\}Z
results would yield 153.64 kcal/mol as the SCF limit, 6.3 kcal/mol
less than the AV\{T,Q,5\}Z+2d1f limit employed in W2 theory. The
AV\{D,T,Q\}Z+2d limit, on the other hand, if fairly close to the
latter at 159.7 kcal/mol. (Our best SCF limit is 159.90 kcal/mol, of
which the extrapolation accounts for 0.15 kcal/mol.)

This type of variability is almost completely absent for the
correlation energy, where AV$n$Z, AV$n$Z+2d and AV$n$Z+2d1f largely
yield the same answers. Following the W2 protocol, the 
CCSD correlation energy is extrapolated using the $A+B/l^3$ extrapolation
formula of Halkier et al.\cite{Hal98} to CCSD/AV\{Q,5\}Z+2d1f energies
(for which $l$=\{4,5\}). (For a fairly comprehensive review of theoretical
and empirical arguments
in favor of this type of extrapolation, see Ref.\cite{w1} and references
therein.) We thus obtain 165.94 kcal/mol as our best
estimate for the CCSD correlation contribution to TAE. It should be
noted that the extrapolation accounts for 3.2 kcal/mol of this amount:
basis set convergence is indeed quite slow. We note that the largest
direct CCSD calculation took a solid two weeks of CPU time
on the DEC Alpha --- a conventional calculation would have required
about 60 GB of temporary disk space, as well as a much higher I/O 
bandwidth if a reasonable wall time to CPU time ratio were to be attained.

As a general rule, the (T) contribution converges much more rapidly with
basis set (besides being smaller to begin with) and therefore, we
were able to dispense entirely with the CCSD(T)/AV5Z+2d1f calculation.
From CCSD(T)/AV\{T,Q\}+2d1f results and the $A+B/l^3$ formula, we
obtain a basis set limit for the (T) contribution of 20.17 kcal/mol, in
which the extrapolation accounts for 0.57 kcal/mol. Together with the
CCSD results, this adds up to a valence correlation contribution
to TAE[SO$_3$] of 186.11 kcal/mol, of which 3.75 kcal/mol is covered by
extrapolations.  

The inner-shell correlation contribution (Table 2) at the CCSD(T) level using 
the Martin-Taylor\cite{hf} core-correlation basis set, was found to be 
0.89 kcal/mol with the Martin-Taylor\cite{hf} core correlation basis
set, and 0.96 kcal/mol with the somewhat more compact MTsmall basis
set used in W2 theory\cite{w1}. Bauschlicher and Ricca\cite{Bau98} found
that basis set superposition error significantly affects the inner-shell
correlation contribution in SO$_2$. It was evaluated here using the
site-site counterpoise method\cite{Wel83} ; we thus found 
counterpoise-corrected core correlation contributions of 0.73
kcal/mol with the Martin-Taylor and 0.68 kcal/mol with the MTsmall
basis sets.

Scalar relativistic effects were obtained as expectation values of the
mass-velocity and Darwin operators\cite{Mar83} for the ACPF (averaged
coupled pair functional\cite{Gda88}) wavefunction. Their effect on 
the computed TAE (with either core correlation basis set) is
-1.71 kcal/mol, comparable to the -1.88 kcal/mol previously found\cite{sif4} 
for SiF$_4$. Atomic spin-orbit splitting adds another -1.23 kcal/mol to the
result. (These latter two terms together imply a relativistic contribution
of -2.94 kcal/mol, or nearly 1\% of the atomization energy.)

Finally, we obtain a W2 total atomization energy
at the bottom of the well, TAE$_e$, of 344.03 kcal/mol; using the
BSSE-corrected inner shell correlation contribution, this value drops to
343.76 kcal/mol. In combination
with the very accurate ZPVE=7.795 kcal/mol\cite{so3}, we finally
obtain, at absolute zero,
TAE$_0$=336.17 kcal/mol without, and 335.96 kcal/mol with, BSSE correction
on the core correlation contribution. This latter value is in perfect
agreement with the experimental TAE$_0$=335.92$\pm$0.19 listed 
in the Gurvich compilation\cite{Gur89}. We thus see once more the importance
of including BSSE corrections for the inner-shell correlation part of
TAE: it should be noted that while the inner-shell contribution to TAE
is small, the S($2s,2p$);O($1s$) absolute 
correlation energy is comparable with the
valence correlation energy in SO$_3$. BSSE on the valence contribution
is much less of an issue since the basis sets used for valence correlation
are much more saturated to begin with, and furthermore the valence 
correlation energy is being extrapolated to the infinite-basis 
limit where it should vanish by definition.

The performance of more approximate computational thermochemistry schemes
is of some interest here (Table 3). G1 theory\cite{g1} is in error 
by no less than -11.4 kcal/mol, which goes down to -6.9 kcal/mol for G2
theory\cite{g2} and -5.45 kcal/mol for G3 theory\cite{g3}. (Only the latter includes
spin-orbit splitting as part of the protocol: none of these methods 
consider scalar relativistic effects.) G2(MP2) performs relatively well
as a result of error compensation (-2.4 kcal/mol). The CBS-Q\cite{cbs94} scheme 
underestimates the true binding energy by only 1 kcal/mol, while 
CBS-QB3\cite{cbs-qb3} is only 0.2 kcal/mol above experiment. It should be
noted that neither CBS-Q nor CBS-QB3 include relativistic effects of any
kind as part of the standard protocol; therefore the excellent performance
put in by these methods is to a large extent thanks to error compensation.
Finally, the W1 theory of Martin and de Oliveira --- which yields a 
mean absolute error of about 0.3 kcal/mol for a wide variety of 
compounds --- has an error in TAE$_0$[SO$_3$] of -1.13 kcal/mol. (W1 theory
includes both scalar relativistic and spin-orbit contributions.)

The 
largest calculations involved in the W1 protocol are CCSD/AVQZ+2d1f
and CCSD(T)/AVTZ+2d1f, which is still rather more demanding than the
steps in any of the G$n$ or CBS methods. Hence this performance is rather
disappointing --- a failure of W1 theory was also noted for SO$_2$
in the original paper\cite{w1}. Balance considerations\cite{so2}
may lead us to wonder whether an AVTZ+2d1f basis set is not rather
top-heavy on inner polarization functions. Using the AV$n$Z+2d
series favored by Bauschlicher and coworkers (e.g.\cite{Bau98}) indeed
reduces the discrepancy with experiment by
0.55 kcal/mol (of which 0.20 kcal/mol in the SCF part). The alternative
sequence \{AVDZ+2d,AVTZ+2d,AVQZ+2d1f\} yields even better agreement
with experiment (and the more rigorous calculations): in fact, the final
value thus obtained falls within the experimental error bar. Particularly
encouraging is the fact that the predicted SCF limit is now within 0.04
kcal/mol of our best estimate. Preliminary calculations on other second-row
systems suggest that this procedure, which we will label W1$'$ theory,
may be preferable over standard W1 theory for second-row systems with
strong inner shell polarization. 
(The two variants are equivalent for first-row compounds.) 

As a test, we have taken three molecules for which W1 yields fairly
large errors (CS, SO, and SO$_2$) and repeated the calculation using
W1$'$ theory. Deviations from experiment drop from 
-0.92, -0.62, and -1.01 kcal/mol, respectively, to
-0.56, -0.32, and -0.02 kcal/mol, respectively, which is not 
qualitatively different from the vastly more expensive W2 calculations
which yielded\cite{w1} deviations of -0.51, +0.02, and +0.23 kcal/mol
for these molecules. We conclude that W1$'$ theory indeed represents
an improvement, and recommend it for future work on second-row systems
instead of W1 theory.

\section{Conclusions}

Benchmark ab initio calculations using direct coupled cluster methods
predict the total atomization energy at 0 K of SO$_3$ to be 335.96 
(observed 335.92$\pm$0.19) kcal/mol. The computed results includes
extrapolation to the basis set limit (3.75 kcal/mol), relativistic effects
(-2.94 kcal/mol), inner-shell correlation (0.68 kcal/mol after BSSE 
correction), and anharmonic zero-point energy (7.755 kcal/mol). 
Inner polarization functions make very large (40 kcal/mol with
$spd$, 10 kcal/mol with $spdfg$ basis sets) contributions to the SCF
part of the binding energy. The molecule presents an unusual hurdle
for less computationally intensive theoretical thermochemistry methods
and is proposed as a benchmark for them.
A slight modification of W1 theory\cite{w1} is proposed which appears to
result in improved performance for second-row systems with strong inner-shell
polarization effects.

\acknowledgments

JM is a Yigal Allon Fellow,
an Honorary Research Associate (``Onderzoeksleider
in eremandaat'') of the
National Science Foundation of Belgium (NFWO/FNRS), 
and the incumbent of the Helen and Milton A. Kimmelman Career 
Development Chair. 
He thanks Prof. Peter J. Knowles (Birmingham
University, UK) for assistance with the installation of 
the direct coupled cluster code, and Dr. Charles W. Bauschlicher Jr.
(NASA Ames Research Center, Moffett Field, CA) for critical reading
of the manuscript prior to submission.
This research was supported by the Minerva Foundation, Munich, Germany.

\begin{table} 
\caption{Convergence behavior of SCF and valence correlation energy of
SO$_3$ (kcal/mol)}
\squeezetable
\begin{tabular}{ldddd}
 & regular & +d & +2d & +2d1f \\
\hline
SCF\\
\hline
AVDZ        &   99.83  & 133.11 & 140.32 & [140.32] \\
AVTZ        &  140.75  & 152.17 & 156.40 &  157.54 \\
AVQZ        &  150.55  & 157.13 & 159.14 &  159.30 \\
AV5Z        &          &        &        &  159.75 \\
Feller(DTQ)$^a$ &  153.64  & 158.89 & 159.70 &  159.50 \\
Feller(TQ5)$^a$ &          &        &        &  159.90 \\
(b)         &          &        &        &  159.93 \\
\hline
CCSD\\
\hline
AVDZ        &  141.21  & 141.10 & 141.49 & [141.49] \\
AVTZ        &  150.93  & 151.41 & 151.46 &  152.19  \\
AVQZ        &  159.36  & 159.60 & 159.67 &  159.74 \\
AV5Z        &          &        &        &  162.76 \\ 
W1 type limit$^c$  &  164.90  & 164.98 & 165.04 &  164.69 \\
(d)         &          &        &        &  165.16 \\
W2 type limit$^e$  &          &        &        &  165.94 \\
\hline
(T)\\
\hline
AVDZ        &   14.89  &  14.97 &  15.01 & [15.01] \\
AVTZ        &   18.74  &  18.76 &  18.76 & 18.82   \\
AVQZ        &          &        &  19.59 & 19.60   \\
W1 type limit$^c$      &   20.17  &  20.16 &  20.16 & 20.24  \\
(d)         &          &        &        & 20.24 \\
W2 type limit$^e$      &          &        &        & 20.17 \\
\end{tabular}

(a) Geometric extrapolation\cite{Fel92} $A+B/C^n$ 
from three points indicated in parentheses

(b) from AVDZ+2d, AVTZ+2d, AVQZ+2d1f series (see text)

(c) two-point extrapolation $A+B/l^{3.22}$ from \{AVTZ,AVQZ\} points for
CCSD, and \{AVDZ,AVTZ\} for the (T) contribution. The empirical exponent
3.22 was determined in Ref.\cite{w1} to maximize agreement with 
more rigorous calculations

(d) from AVDZ+2d, AVTZ+2d series (see text)

(e) two-point extrapolation\cite{Hal98} $A+B/l^3$ from \{AVQZ,AV5Z\} points for 
CCSD, and \{AVTZ,AVQZ\} for the (T) contribution. 

\end{table}

\clearpage

\begin{table}
\caption{Computed and observed total atomization energy of SO$_3$ (kcal/mol) at 0 K}
\squeezetable
\begin{tabular}{lddddd}
                        &  W1   & W1'  & W2  & W2 (a) & best \\
\hline
SCF                     &159.50  & 159.93&159.90 & 150.90&159.90\\
Valence correlation     &184.93  & 185.40&186.11 & 186.11&186.11\\
Inner-shell correlation & +0.96  & +0.96 & +0.96 & +0.68 & +0.73 \\
Scalar relativistic     & -1.70  & -1.70 & -1.70 & -1.70 & -1.71 \\
Atomic spin-orbit       & -1.23  & -1.23 & -1.23 & -1.23 & -1.23 \\
TAE$_e$                 &342.46  &343.44 &344.03 &343.75 &343.79\\
Zero-point energy       &  7.60$^b$  &  7.60$^b$ &  7.79 &  7.79 &  7.79\\
TAE$_0$                 &334.86  &335.77 &336.24 &335.96 &336.00\\
Experiment\cite{Gur89}  &        &       &       &       &335.92$\pm$0.19\\
\end{tabular}

(a) with BSSE correction to core correlation (see text)

(b) Following W1 protocol, from  B3LYP/VTZ+1\cite{Bec93} harmonic frequencies
scaled by 0.985.\cite{w1}

\end{table}

\begin{table}
\caption{Comparison of computed and observed atomization energies (kcal/mol)
for SO$_3$ using different computational thermochemistry protocols}
\squeezetable
\begin{tabular}{lddd}
              & $TAE_e$ & $TAE_0$  & error\\
\hline
G1           & 332.24 & 324.52 & -11.40\\
G2           & 336.72 & 329.00 & -6.92\\
G2MP2        & 341.30 & 333.58 & -2.34\\
G3           & 338.19 & 330.47 & -5.45\\
CBS-Q        & 342.79 & 334.88 & -1.04\\
CBS-QB3      & 343.62 & 336.13 &  0.21\\
W1           & 342.46 & 334.86 & -1.06\\
W1'          & 343.37 & 335.77 & -0.15\\
W2           & 344.04 & 336.24 &  0.32\\
W2     (a)   & 343.76 & 335.96 &  0.04\\
Experiment   &        & 335.92$\pm$0.19\\
\end{tabular}

(a) including BSSE correction on the inner-shell correlation
contribution (see text)

\end{table}

\end{document}